\documentclass[lp]{elsarticle}

\usepackage{graphicx}
\usepackage{amsfonts}
\usepackage{amsmath}
\usepackage{hyperref}
\usepackage{float}
\usepackage{relsize}

\newcommand{\be}{\begin{equation}}			
\newcommand{\ee}[1]{\label{#1} \end{equation}}		
\newcommand{\ba}{\begin{eqnarray}}			
\newcommand{\ea}[1]{\label{#1} \end{eqnarray}}		
\newcommand{\nl}{\nonumber \\}				

\newcommand{\td}[2]{ \frac{{\rm d\,} #1}{{\rm d\,} #2}}				

\begin{document}



\title{
{\bf Ideal gas provides q-entropy}
}


\author[els]{T.\ S.\ Bir\'o}
\ead{Biro.Tamas@wigner.mta.hu}
\address[els]{Wigner Research Centre for Physics of the HAS, 
H-1525 Budapest, P.O.Box 49, Hungary}



\begin{abstract}
A mathematical procedure is suggested to obtain deformed entropy formulas 
of type $K(S_K)=\sum{P_i K(-\ln P_i)}$,
by requiring zero mutual $K(S_K)$-information between a finite subsystem and 
a finite reservoir. The use of this method is first demonstrated on the ideal
gas equation of state with finite constant heat capacity, $C$, where it delivers the
R\'enyi and Tsallis formulas.  A novel interpretation of the $q^*=2-q$ duality 
arises from the comparison of canonical subsystem and total microcanonical partition approaches.
In the sequel a new, generalized deformed entropy formula is constructed for the 
linear $C(S)=C_0+C_1S$ relation. 
\end{abstract}


\begin{keyword}
q-entropy \sep mutual information \sep finite heat reservoir
\end{keyword}

\maketitle

\section{Introduction \label{SEC:INTRO} }


Besides the familiar logarithmic Boltzmann-Gibbs-Shannon(-Planck) entropy
formula, valid for most ergodic systems in the thermodynamical (infinite number
of degrees of freedom) limit, generalizations of the entropy -- probability
relation have occured several times. These relations usually contain
one or more parameters besides the Boltzmann constant, $k_B$. 
In this paper we refer to them as ''q-entropy'', summarizing in $q$
these extra parameters. The physical origin and quantitative determination
of such parameters from microscopical theories are longstanding but
still actual and intriguing questions.

According to the view of the present author its universal origin is best
understood by the study of finite reservoir effects, usually neglected
in the classical thermodynamical limit. In particular, large but finite
reservoirs may show effects relying on ratios, such as the
entropy to heat capacity, $S/C$ ratio. Thermal models are 
unexpectedly successfull when applied to hadron energy spectra
arising in high energy experiments, although an infinite heat reservoir
is not to be found. Theoretical calculations considering a
finite reservoir are necessary for a more advanced statistical approach.

The thermodynamical study of finite systems is also motivated by the 
need for physical interpretation and modeling
of the parameters occurring in deformed entropy formulas.
Most known of such relations are the one-parameter ($q$) tagged
R\'enyi- and Tsallis-formula
\cite{Renyi:1970,Jizba:2003gx,Bialas:2008zz,Tsallis:1988,Tsallis:1998,Tsallis:1999,TsallisBook}.

In this paper we point out how deformed entropy formulas
emerge from requiring zero mutual information between finite parts
of some ideal thermodynamical systems. 
First we concentrate on the classical ideal gas\cite{Naudts} with
finite constant heat capacity \cite{Almeida:2001,Rybczynski:2004gs}, 
then we consider possible generalizations including more complex systems, 
having a linear heat capacity - entropy relation.
We hope that once an understanding can be deepened for the ideal gas,
considered to be thermodynamically the simplest possible system, one may
hope to gain orientation for describing the thermal behavior of more
complex systems. In fact finite reservoir effects for a wide class of
scaling Hamiltonians have been shown to lead to 
energy distributions of a cut power-law form \cite{ScalingHam}. 
This distribution is formally a canonical energy distribution to
the one parameter q-entropies of R\'enyi and Tsallis.

The ideal gas may be viewed from several viewpoints depending on which of its
properties is taken as the defining one. In high energy physics it is widespread
to formulate that the ideal gas has no interaction. This is not precise,
since even elastic (kinetic energy conserving but momentum changing)
collisions represent short term, but violent interactions.
Another, macroscopical, viewpoint considers the Boyle-Mariotte law,
$pV=NT$ (with $p$ pressure in $V$ volume for $N$ atoms at $T$ absolute
temperature measured in units where the Boltzmann constant is set to
one, $k_B=1$), as defining what an ideal gas is. This relation, however,
is valid for any equation of state (eos) in the form
$S=N \ln V + A(E,N)$, considering a too wide class of exotic, 
energetically complex systems.
From energetic viewpoint the constant heat capacity systems may be considered
as ideal gases. This definition, on the other hand, would exclude among others
black body radiation and a relativistic quark-gluon plasma, where
the number of particles is not fixed. To overcome this restriction
in the last section of this article we deal with
systems with entropy-dependent heat capacity, $C(S)$.

Finally it occurs as an {\em ideal} property if a part of a system is similar
to the whole. This means a mathematical factorization between subsystem and
reservoir, even in the case of constraints connecting their states -
alike the most known fixed total energy constraint. In fact the mutual information,
which can be gained about a part of a system by observing the state of its
complement, classically coincides with the entropic difference
\be
 I_{12}=S_1+S_2-S_{12}
\ee{MUTUAL_ENTROPIC}
if denoting the corresponding (logarithmic) entropies by $S_i$ with
$i$ referring to the parts with indices $1$ and $2$ of the total system
with index $12$. For the fixed total energy problem it reads as
\be
I_{12} = S_1(E_1) + S_2(E_2) - S_{12}(E_1+E_2) \ne 0.
\ee{MUTUAL_FIXED_E}
This quantity is not zero, even for the classical ideal gas, as it will be
shown below. This fact will be called {\em non-additivity} of the entropy $S$
throughout the present paper. This is not equivalent with the homogeneity
of the ideal gas equation of state in the properly normalized treatment;
where $S(E,V,N)=Ns(E/N,V/N)$. Such expressions are additive, {\em provided}
the parts have equal densities, $E_1/N_1=E_2/N_2=e$ and $V_1/N_1=V_2/N_2=v$.
In this case $(E_1+E_2)/(N_1+N_2)=e$ and $(V_1+V_2)/(N_1+N_2)=v$
also holds. We aim at a zero mutual information description between finite
parts even for unequal densities of the thermodynamical extensives.

In order to achieve factorization we seek for a mathematical description by obtaining
a $K(S)$ relation, called ''deformed'' entropy, which provides zero mutual
information,
\be
I_{12}^{(K)} = K_1(S_1(E_1)) + K_2(S_2(E_2)) - K_{12}(S_{12}(E_1+E_2)) = 0.
\ee{ZERO_K_MUTUAL}
It is not a priori trivial whether such $K(S)$ functions exist, and how
universal, e.g. independent of the thermodynamical intensives, they can be
\cite{AsympRules,BiroBook,ZerothLaw}.

By the construction of $K(S)$ with zero mutual K-information, additivity in the
above sense is achieved. Having an additive $K(S)$ the general ensemble-formula
can be derived as follows. The additivity of two systems extends to the additivity
of an arbitrary number (by the basic recurrence property of ordinary summation):
\be
\sum_i K_i\left(S_i(E_i)\right) = K_{{\rm tot}}\left(S_{{\rm tot}}\left(\sum_i E_i\right) \, \right).
\ee{N_ADDONS}
We note that this property is equivalent to the requirement that the generic
$K(S)$ relation is a linear function of the energy; \hbox{$\partial^2 K(S(E)) / \partial E^2 = 0$.}
Applying eq.(\ref{N_ADDONS}) to $n$ equal systems one obtains
\be
n K_1\left(S_1(E_1)\right) = K_n \left(S_n(nE_1)\right).
\ee{EQUAL_ADDONS}
In a more general setting the system with energy $E_i$ and equation of state $S_i(E_i)$
occurs $n_i$ times, leading to
\be
\sum_i n_i K_i \left(S_i(E_i) \right) = K_{{\rm tot}}\left( S_{{\rm tot}} \left(\sum_i n_i E_i \right) \, \right).
\ee{DEGER_ADDONS}
Finally defining $n=\sum_j n_j$ and the relative occurrence frequency 
of energy $E_i$ in the ensemble as $P_i=n_i/n$ (these ratios being between zero and
one and summing up to one), we rewrite eq.(\ref{DEGER_ADDONS}) as
\be
 n \sum_i P_i K_i \left( S_i(E_i) \right) =
 K_{{\rm tot}} \left( S_{{\rm tot}} \left(n \sum_i P_iE_i \right) \, \right).
\ee{REWR_ADDONS}
Applying eq.(\ref{EQUAL_ADDONS}) to the right hand side and dividing both sides by $n$
finally we achieve
\be
 \sum_i P_i K_i \left( S_i(E_i) \right) =
 K_{{\rm tot/n}} \left(S_{{\rm tot/n}} \left(\sum_i P_i E_i \right) \, \right).
\ee{FINAL_ADDONS}
Utilizing now the Boltzmann-conjecture (based on the Stirling formula for large factorials)
we replace $S_i(E_i)=-\ln P_i$ to arrive at
\be
 K(S_K) = \sum_i P_i K_i\left(-\ln P_i \right)
\ee{K_PROBA}
where the summation is meant over all possible states, the $i$-th
one with energy $E_i$. Here the quantity $K(S_K)$ is an ensemble-averaged deformed entropy,
it is additive. The deformed ''average''
$S_K$ is the entropy in the Boltzmannian sense, but nothing warrants
its additivity. As the number of observations, $n_i$ approaches
infinity, the relative occurrence frequencies, $P_i$ come close to the
thermodynamic probability.
We note that if the Boltzmannian entropy were additive (at least to some approximation),
then $K_i(S_i)=S_i$ is (at least close to) the identity function. For this case the
classical formula emerges:
\be
S_{{\rm id}} = S_{{\rm Gibbs}} = \sum_i P_i (-\ln P_i).
\ee{SGIBBS}
The procedure described above can be done in two dual ways, 
$P_1$ and $K(S_1)$ once can be associated to 
the $S_1(E_1)$ term, once another $P_1^*$ and $K^*(S_1)$ to the 
$\Delta S(E_1)=S_{12}(E)-S_2(E-E_1)$ expression.
We obtain
\be
 K(S_K) \, = \,  \sum_i P_i K(-\ln P_i)
\ee{LENTR}
for the thermodynamical and
\be
 K^*(\Delta S_K) \, = \, \sum_i P_i^* K^*(-\ln P_i^*)
\ee{STAR_KENTROPY}
for the canonical probabilities in a finite subsystem with finite reservoir.
We contemplate on the different physical content behind these
respective formulations in the next section.


\section{Microcanonical Equation of State for an Ideal Gas}

In the statistical mechanics the ideal gas is viewed as a mechanical system
of $N$ particles completely described by their kinetic energy. 
The energy-momentum relation, $E(\vec{p})$, connects 
the classical uniform phase space distribution to a state density $\rho(E)=d\Gamma/dE$.
In particularly simple cases, like the non-relativistic
kinetic energy of massive particles, $E=\vec{p}^2/2m$,
the state density has the form $\rho(E)\propto V E^f$ with $f=D/2$
for $D$-dimensional momentum vectors, $\vec{p}$.

Fixing the total energy with
a uniform occupation of a spherical shell in the $2ND$-dimensional phase space,
the dependence on the $D$-dimensional volume, $V$ and on the total energy, $E$ 
is easy to obtain for a fixed particle number, $N$:
\be
W_N(E)\!=\!e^{S(E)}= \frac{1}{N!} \!\int\!\prod_{j=1}^N\limits\!d\Gamma_{\!j} \delta\!\!\left(\!E-\sum_{i=1}^N\limits E_i\!\right)  
   \!\propto\! V^{N} E^{fN-1}
\ee{SHELL_VOLUME_SCALING}
with $d\Gamma = dV d^Dp / (2\pi\hbar)^D$.
Boltzmann has suggested that the a priori equally probable states are realized
with a relative frequency of $P=1/W=e^{-S(E)}$ in a huge number of observations.
$1/W$ is traditionally called ''thermodynamical probability''.
From here the entropy, $S=\ln W$ \cite{KardarBook} 
(with unit Boltzmann constant $k_B=1$) is given as
\be
 S(E,V,N) = S_0(N) + N \ln V + C_N \ln E.
\ee{ENTROPY_MICROCAN_SHELL}
Such an equation of state has an energy and volume independent heat capacity,
$C_N=C_{V,N}=fN-1$, constant for a fixed particle number, $N$ \cite{KardarBook}.
From now on we simplify the discussion to fixed volume and particle number,
dealing with the equation of state in form of a single $S(E)$ relation.
It also means that when considering entropy non-additivity at energy-additivity
we have the same $N$ and $V$ values in mind.
It is particularly interesting for high energy physics, where sometimes
the particle number and the volume do not play a role among the sensible
control parameters.
Primes denote then derivatives with respect to the corresponding single argument.

We consider a class of equations of state with 
{\em finite constant heat capacity}. Using the definition,
$C = dE/dT$, one considers
\be
\frac{1}{C} = \td{T}{E} = \td{}{E} \left(\frac{1}{S'(E)} \right)= - \frac{S''(E)}{S'(E)^2}.
\ee{CONSTANT_HEAT_CAPACITY}
With an energy independent value $1/C=1/C_0$ 
the differential equation can be solved for $S(E)$.
The first integral becomes the temperature
\be
T = \frac{1}{S'(E)} = T_0 + \frac{1}{C_0} E,
\ee{FIRST_INTEGRAL_CON_HEAT_CAP}
with $T_0$ being here an integration constant. 
Its physical meaning is associated to the minimal energy, $E_0=C_0T_0$, belonging to $W(E_0)=1$.
The ideal gas temperature is in general a linear function of the energy $E$
(calling this relation the ''equipartition relation''). 
The second integration leads to the general microcanonical equation of state with
finite constant heat capacity:
\be
 S(E) = C_0 \ln \left(1 + \frac{E}{C_0T_0} \right)  + S_0.
\ee{MICRO_EOS_C} 
Counting the energy $E$ relative to the state with minimal classical energy, $E=0$, 
we require $S(0)=0$ according to the third law of thermodynamics. 
This fully defines an ideal gas. 
The coefficients $C_0$ and $T_0$ may differ from problem to problem.
We note, however, that $T_0$ is never exactly zero in real systems, as long as the entropy
$S(E)$ is non-negative. The Seckur-Tetrode relation is a quantum-corrected, but
high energy limit approximation to this result.

The exponentiated entropy upon eq.(\ref{MICRO_EOS_C}) with $S_0=0$
\be
 W = e^{S(E)} = \left( 1 + \frac{E}{C_0T_0} \right)^{C_0}
\ee{expS}
delivers the thermodynamical probability factor of finding a sytem with energy $E$
by a uniformly random search in the phase space. It is normalized to $W(0)=1$.
For an {\em isolated} system having energy $E_1$
this probability factor is \cite{Naudts}
\be
P_1 =  \frac{1}{W_1(E_1)} = e^{-S_1(E_1)} =  \left( 1 + \frac{E_1}{C_0T_0} \right)^{-C_0}.
\ee{ALONE}
On the other hand the conditional probability of a {\em subsystem} - indexed by $1$ -
having energy $E_1$ while the total energy is fixed to $E$ becomes:
\be
P^*_1 := \frac{W_2(E-E_1)}{W_{12}(E)} \propto  \left(1 - \frac{E_1}{C_{02}T} \right)^{C_{02}}
\ee{P_SMICRO}
with $T=T_{02}+E/C_{02}$ and $C_{02}$ being the heat capacity of the reservoir (system 2).
Note that - after some algebra - in the denominator $T$ occurs in place of $T_{02}$.
This factor is referred to in the canonical treatment of thermodynamical systems.
In the infinite $C_{02}$ limit  an exponential distribution emerges,
\be
 P^*_1 \longrightarrow  e^{-E_1/T}.
\ee{IDEALGASCANO}
This result is frequently quoted as the ''equivalence of the canonical
and microcanonical statistics in the thermodynamical limit''. It, however, does not
hold for all arbitrary ratios of fluctuations in arbitrary systems 
\cite{VolumeFluct,BoseFermiFluct,ExactChargeFluct,MicroFluct,CanonFluct,Torrieri2010,Wilk:TempVolFluct}.
The ideal gas with a finite heat reservoir has been recently studied and 
showed to lead to a density of states
in the subsystem which interpolates between microcanonical and canonical
distributions as the heat capacity of the reservoir part grows from zero
to infinity \cite{Campisi}. This underlines the importance of the cut power-law
type energy distribution (\ref{P_SMICRO}) from another viewpoint.

In a finite ideal gas the canonical probability for a subsystem with a given energy
and the Boltzmannian thermodynamical probability vastly differ,
\be
P^*_1 =e^{S_2(E-E_1)-S_{12}(E)}  \ne e^{-S_1(E_1)} = P_1, 
\ee{MAXPR}
basically due to the {\em non-additivity} (non-zero mutual information) 
of the microcanonical entropy.  While the energy adds up, the ideal gas entropy does not.

It is a relevant question which probability influences experimental observations.
Regarding a long term observation of $N$ particles with total energy $E$, a
single particle has energy $E_1$ with the frequency $P^*_1$.
Here, however, one supposes that the {\em observed particle remains in the system}
for infinitely many consecutive observations. If the {\em particle is taken out}
after each observation, then the energy and number of particles both diminish
during detection. Then to observe first an energy $E_1$ out of $E$ and then
an energy $E_2$ out of $E-E_1$ etc. delivers
\be
 P^*_1 P^*_2 \ldots P^*_N = 
\nonumber
\ee{EMPTY1}
\be
= \frac{W_{N-1}(E-E_1)}{W_N(E)}\frac{W_{N-2}(E-E_1-E_2)}{W_{N-1}(E-E_1)} \ldots 
   \frac{W_0(0)}{W_1(E-\ldots -E_{N-1})} 
\nonumber
\ee{EMPTY2}
\be
  =  \frac{1}{W_N(E)} = P_N(E), 
\ee{INFINITE_PRODUCT}
due to $W_0(0)=1$ and the steady cancellation of $W$-factors in the chain of ratios.
This points out the physical meaning of the thermodynamical probability, $1/W$.
So the observation of any partition of the initial energy $E$ by a decay
is equally probable.

\section{\lowercase{q}-Entropy of the Ideal Gas}

A possible improvement towards mathematical beauty 
is achieved if factorization with zero mutual information is performed. 
This procedure can be applied both to the thermodynamical (\ref{ALONE}) 
and conditional (\ref{P_SMICRO}) probabilities.
In this sense the part of an ideal gas becomes itself a similar ideal gas.
Generalizing Boltzmann's idea one seeks for a probability, based on
a given function of the original entropy, $K(S)$.
Considering 
\be
 P_K(E) = e^{-K(S(E))},
\ee{K_PROB}
one uses an additive function of the original entropy, a ''formal logarithm'' \cite{AsympRules,BiroBook}.
Since we assume additivity of the energy, {\em this is a linear expression of $E$}.
By using the functional form inverting the ideal gas equation of state with finite constant
heat capacity, eq.(\ref{MICRO_EOS_C})  exactly, we have 
\be
 K(S) = \lambda E + \mu = \lambda C_0T_0\left( e^{S/C_0}-1 \right) + \mu. 
\ee{ALMOST_TSALLIS_ENTROPY}
Satisfying the natural requirement of $K(S)\approx S$ for small $S$ one fixes
$K(0)=0$ and $K'(0)=1$. This determines the linearity parameters as
$\mu=0$ and $\lambda=1/T_0$. We get
\be
 K(S) = C_0\left( e^{S/C_0}-1 \right). 
\ee{TSALLIS_ENTROPY}
Indeed for infinite heat capacity systems, $C_0\rightarrow\infty$ in the above formula
$K(S)\rightarrow S$, provided that $S/C_0$ becomes arbitrarily small.
It is important to realize that this result is {\em independent of the
integration constant} $T_0$.
Using this construction
\be
 P_K^*(E_1) := \frac{e^{K(S(E-E_1))}}{e^{K(S(E))}} = e^{-K(S(E_1))} = P_K(E_1).
\ee{L_MICRO}
Obviously $ K(S_1(E_1)) + K(S_2(E-E_1)) = K(S_{12}(E))$
by using the same $T_0$,
i.e. the deformed entropy for the finite ideal gas is {\em additive} 
and the separation with zero mutual information is performed.

We can apply this procedure for the canonical probability in another way.
Seeking for an additive function of  
{$\Delta S = S_2(E-E_1)-S(E)$,} denoted by $K^*(\Delta S)$,
we consider $\Delta S=-\ln P^*$ and set the subsystem-energy linearity by
\be
K^*(-\ln P^*) = \lambda E_1 + \mu.
\ee{ALMOST_TSALLIS_STAR}
Inverting eq.(\ref{P_SMICRO}) we obtain
\be
 K^*(\Delta S) = \lambda C_0T \left(1-e^{-\Delta S/C_0} \right) + \mu.
\ee{ALMOST_ALMOST_STAR}
With the natural conditions $K^*(0)=0$ and $(K^*)'(0)=1$ finally we arrive at
\be
 K^*(\Delta S) = C_0 \left(1-e^{-\Delta S/C_0} \right).
\ee{TSALLIS_ENTROPY_STAR}
It is noteworthy that this result is independent of the temperature, $T$.
The factorizing 'star'-quantity becomes
\be
P_{K^*}(E_1) := e^{-K^*(-\ln P^*(E_1))} = e^{-E_1/T}.
\ee{STAR_FACTOR_QUANTITY}


Now we consider the additive entropy based on the thermodynamical probability.
Applying our result (\ref{TSALLIS_ENTROPY}) we arrive at 
\be
  K(S_K) = C_0 \sum_i P_i (e^{(-\ln P_i)/C_0}-1). 
\ee{TK_FORMULA}
By using $q=1-1/C_0$ (with $C_0$ being the heat capacity of the isolated system) 
this expression takes the form of Tsallis q-entropy:
\be
S_{{\rm Tsallis}} = K(S_K) = \frac{1}{1-q} \sum_i \left(P_i^q-P_i \right).
\ee{Tq_FORMULA}
It is noteworthy that $S_K$ becomes now the R\'enyi formula:
\be
S_{{\rm Renyi}} = S_K = \frac{1}{1-q} \ln \sum_i P_i^q.
\ee{RS_FORMULA}
Next we consider the additive entropy based on the canonical probability.
Based on (\ref{TSALLIS_ENTROPY_STAR}) we obtain 
\be
  K^*(\Delta S) = C_0 \sum_i P^*_i (1-e^{(\ln P^*_i)/C_0}) 
\ee{TK_STAR_FORMULA}
with $\Delta S=S_{12}(E)-S_2(E-E_1)$.
Using now $q^*=1+1/C_0$ (with $C_0$ being the heat capacity of the reservoir) 
this expression also takes the form of a Tsallis
formula, but with a dual parameter $q^*=2-q$:
\be
S_{{\rm Tsallis}} = K^*(\Delta S) = \frac{1}{1-q^*} \sum_i \left((P_i^*)^{\: q^*}-P^*_i \right).
\ee{TqSTAR_FORMULA}
Again the original entropy factor, $\Delta S=S_{12}(E)-S_2(E-E_1)$ is given as a R\'enyi formula:
\be
S_{{\rm Renyi}} =  \Delta S = \frac{1}{1-q^*} \ln \sum_i (P_i^*)^{\: q^*}.
\ee{RS_STAR_FORMULA}
This relation between thermodynamical and canonical probabilities for an ideal gas
show unexpectedly a new facet of the ''$q$-duality'' \cite{TsallisBook}.

\section{Formally Canonical Treatment}

The factorizing quantities, $P_K$ and $P_{K^*}$, are exponential functions
of the additive energy variables, they formally look ''canonical'', 
c.f. eq.(\ref{STAR_FACTOR_QUANTITY}).
However, please note that $T_0 \ne T$. For positive heat capacity (''normal'') physical systems
$T > T_0$. This is the basic reason why cut power-law fits to
observed spectra give a smaller slope parameter extrapolated to zero energy than direct fits to an exponential
function
\cite{Biro:2008er,Biro:2008hz,Urmossy:2011xk,Wilk:2009aa,Wilk:JPG39,Wilk:2012,CMSfit,Cleymans:2008mt,CleymansWorku,RHIC_Participant}.
Supporting these ensemble-entropy formulas by conditions of probability normalization
and average energy one arrives at the formal {\em K-canonical} principle
\ba
 & & \left(\sum_i P_i K(-\ln P_i) - K(S_K(E)) \right) 
 	 - \alpha \left(\sum_iP_i-1\right)
\nl
& - &	  \beta \left( \sum_iP_iE_i - \langle E \rangle \right) \, = \, {\rm max.}
\ea{LCANON}
for a generic $P_i$ distribution.
This view already includes factorization and is the form compatible with
a temperature definition independent of the subsystem-rest division
\cite{BiroDeriv:2012,VanKazany:2012}.

Thus we arrive at the Tsallis and R\'enyi formulas for the q-entropy
either with $q=1-1/C_0$ or with $q^*=1+1/C_0$ depending on which probability is factorized:
In the former case the isolated systems' thermodynamical probability, in the
second case the canonical finite subsystem - finite reservoir probability.
It is a curious fact that these, in their origin microcanonical probabilities 
can be obtained from a formally canonical treatment by the variational principle (\ref{LCANON}).
The formally canonical distribution, constrained by the average energy and  probability normalization
is given by the stationary point of eq.(\ref{LCANON}) with respect to all $P_i$-s:
\be
 P_i = \left(1+ \frac{1+\alpha + \beta E_i}{C_0-1}\right)^{-C_0}.
\ee{LCANOPROB}
in the case $q=1-1/C_0$ (\ref{ALONE}). Here $C_0$ is the heat capacity of the total system
under consideration.
The canonical single particle energy distribution in an
undisturbed $N$-particle ideal gas is obtained by using $q^*=1+1/C_0$
on the other hand (\ref{P_SMICRO}):
\be
 P^*_i = \left(1- \frac{1+\alpha + \beta E_i}{C_0+1}\right)^{C_0}.
\ee{LCANOPROB_STAR}
Here $C_0$ is the heat capacity of the reservoir.
Usually, for a sensible positive average energy, $\alpha$ and $\beta$ are positive,
so the {\em K-canonical} probability distribution is of one plus energy on a negative
power type, going above the exponential. This exactly reconstructs the thermodynamical
probability for a fixed number ideal gas. The {\em $K^*$-canonical} 
distribution behaves opposite. 


In order to show the strength of this approach we derive some physically important
limits of the finite-canonical probability, $P^*_i$.
One observes that
\be
\sum_i (P^*_i)^{1+1/C_0} = \sum_i \left(1-\frac{1+\alpha+\beta E_i}{C_0+1} \right) P^*_i 
 = 1 - \frac{1+\alpha+\beta U}{C_0+1}.
\ee{ALFA_ENTROPY}
Here $U=\sum_i P^*_i E_i$ is the canonical average energy in case of a finite
heat reservoir. 
Due the  eq.(\ref{TqSTAR_FORMULA}) on the other hand
\be
\sum_i (P^*_i)^{1+1/C_0} =  e^{-\Delta S/C_0},
\ee{WHAT_IS_Pstarq}
so we express $\alpha$ from the above relations as being
\be
\alpha = C_0 - (C_0+1) e^{-\Delta S/C_0} - \beta U.
\ee{ALFA}
Utilizing this result the canonical energy distribution when connected to a finite
heat reservoir is given by
\be
P^*_i = e^{-\Delta S} \left(1-\frac{E_i-U}{(C_0+1)T} \right)^{C_0}.
\ee{CANONP_FINHEAT}
with $1/T=\beta e^{\Delta S/C_0}$.
Applied to the classical ideal gas, $C_0=3N/2-1$ with $N$ particles in the
reservoir. The $N\rightarrow\infty$ limit can be easily obtained by the use of Euler's
formula:
\be
\lim_{N\rightarrow\infty} P^*_i = e^{-\Delta S} e^{-\frac{E_i-U}{T}}.
\ee{INFTY_CANONICAL}
This result agrees with the standard formula.

The opposite limit, $N=0$ is also interesting; in this case the reservoir vanishes.
Observing that in general
\be
\lim_{m\rightarrow 0} \left( 1 - \frac{z}{m} \right)^m = \Theta(-z)
\ee{THETA}
is a representation of the step function, its derivative with respect to $z$
delivers a Dirac delta. Applying to our formula (\ref{CANONP_FINHEAT}) we
obtain the standard microcanonical result
\be
\lim_{N\rightarrow 0} P^*_i  = e^{-\Delta S} \, \delta(E_i-U).
\ee{MICROP}
This interpolating property of the cut power-law energy distribution due to
a finite heat reservoir has also been observed in \cite{Campisi}.



\section{Generalization}

So far we have seen that considering an additive function of a non-additive entropy formula 
while the energy is
kept additive, it {\em can only be} an expression proportional to the energy:
 $K(S(E)) = E/T_0$.
With other words, the second derivative of $K(S(E))$ with respect to the energy
vanishes:
\be
 \td{^2}{E^2} K(S(E)) = 0.
\ee{SEC_DERIV_K}
This relation can be obtained also as a general requirement of having
zero mutual information in the subsystem - reservoir
couple, as it has been shown in section
\ref{SEC:INTRO} and in \cite{BiroDeriv:2012}. 
Applying the chain rule, eq.(\ref{SEC_DERIV_K}) is equivalent to
\be
\frac{K''(S)}{K'(S)} = - \frac{S''(E)}{S'(E)^2} = \frac{1}{C(S)},
\ee{ONE_PER_C_K}
pointing out the central role of the heat capacity of the considered system
by such constructions.
This separable differential equation posseses a general formal solution,
\be
 K'(S) = e^{\int \frac{dS}{C(S)}}
\ee{SOL_K_PRIME}
and
\be
 K(S) = \int e^{\int \frac{dS}{C(S)}} \, dS.
\ee{SOL_K_ITSELF}
Now the first derivative of $K(S(E))$ with respect to the energy, $E$,
is independent of the energy. This makes it possible to satisfy the zeroth law
of thermodynamics universally; the temperature set by a finite reservoir for a subsystem 
is to leading order independent of the energy of the same subsystem. 
This {\em universal thermostat independence} (UTI)
principle\cite{BiroDeriv:2012} allows to define a 
zeroth law compatible temperature in the finite 
microcanonical analysis. Here $C(S)$ is the heat capacity of the total isolated system.

In the canonical treatment on the other hand the physical constitution of the
heat reservoir is decisive and the general requirement is given by
\be
\td{^2}{E_1^2}K^*(S_2(E-E_1)-S(E)) = 0. 
\ee{SEC_DER_STAR}
The separated form of the differential equation becomes
\be
\frac{K^{*\prime\prime}(\Delta S)}{K^{*\prime}(\Delta S)} = 
- \frac{S_2^{\prime\prime}(E-E_1)}{S_2^{\prime}(E-E_1)^2} = \frac{1}{C_2(S_2)}.
\ee{SEPARATED_DIFFEQ}
It is obvious that in this case the heat capacity of the reservoir governs the physics.

For an ideal gas with constant number of particles, $N$, also the heat capacity, $C_{V,N}$,
is constant. After abandoning this fixing
more general connections, $C(S)$, should be taken into account even for ideal gases.
Physically important examples are black body radiation, extensible to models
with bag constant, where $C=3S$, and simple black hole horizon thermodynamics with $C=-2S$.
In these cases $N$ is not a basic thermodynamical variable, in the second case even $V$ is missing.
We present here the corresponding formulas for the linear class of equations of state, 
$C(S)=C_0+C_1S$. In this case the equation of state integrates to
\be
 S(E) = \frac{C_0}{C_1} \left[\left(1+\frac{C_1+1}{C_0T_0}E \right)^{\frac{C_1}{C_1+1}} -1\right].
\ee{CLASS_1_SE}
One obtains
\be
 K^*(\Delta S) = \frac{C}{C_1} \sum_i P^*_i \: \frac{1}{q_1}\left[1-\left(1+\frac{C_1}{C}\ln P^*_i\right)^{q_1}\right]
\ee{CLASS_1_KSTAR_S}
with $q_1=1+1/C_1$ and $C=C_0+C_1S$
for the entropy formula (\ref{STAR_KENTROPY}) associated to the conditional (finite-canonical) probability.
For $C\rightarrow\infty$ the Gibbs formula (\ref{SGIBBS}), for $C_1\rightarrow 0$ the Tsallis formula 
(\ref{TqSTAR_FORMULA}) arises.
An exploration of the potential of these formulas, in particular for $C_0=0$, 
is referred to a future work.

In summary we have shown that a similarity requirement between an ideal gas and its
finite parts points out a certain deformed entropy expression.
For finite constant heat capacity systems the corresponding additive
quantity turns out to be the q-entropy formula promoted by Tsallis. 
For heat capacities being a linear function of the entropy a new deformed
entropy formula arises (eq.\ref{CLASS_1_KSTAR_S}), containing two
deformation parameters.
Further formulas are obtained for a generic $C(S)$ relation by the UTI
principle, see eq.(\ref{SOL_K_ITSELF}).


\vspace{2mm}
{\em Acknowledgment: \quad}
Discussions with Profs. C. Tsallis, G. Torrieri and V. Koch are gratefully acknowledged. 
This work has been supported by Hungarian National Research Fund OTKA grant 
K104260, by the Hungarian-South-African project NIH TET\_10-1\_2011-0061 and ZA-15/2009,
and was also supported by the Helmholtz
International Center for FAIR within the framework of the LOEWE
program 
launched by the State of Hesse.


\end{document}